# Reply to "Limitations of detecting structural changes and time-reversal symmetry breaking in scanning tunneling microscopy experiments"

Yuqing Xing, Seokjin Bae, Stephen D. Wilson, Ziqiang Wang, Rafael M. Fernandes, and Vidya Madhavan

## 1. Summary

In science, cross-examination from different perspectives is always welcome.

In our paper[1] titled "Optical Manipulation of the Charge-Density-Wave State in RbV$_3$Sb$_5$", we demonstrate that the relative intensities of the charge-density-wave (CDW) peaks in the Fourier transform of STM topographies of RbV$_3$Sb$_5$ can be controlled by an electric field induced by laser light. Moreover, we show that this change in the intensities is accompanied by changes in the ratio of Bragg vectors. Additionally, we replicate previous findings on the changes in the CDW state induced by a magnetic field.

In their comment, Candelora et al. assert that, in the absence of artifacts like drift, noise, and tip changes, the charge density wave (CDW) intensity in the 135 compounds would remain unaffected by light and magnetic fields, in line with a previous publication from their group[2]. Before addressing their comments in detail below, we would like to highlight a few key points that cannot be dismissed on the basis of measurement artifacts.

First, science progresses through replication. The switching of the relative CDW intensity order with magnetic field has been reported in STM studies for KV$_3$Sb$_5$ and RbV$_3$Sb$_5$ before[3,4] and after[5] our publication. Our work on RbV$_3$Sb$_5$[1] replicates the original findings on KV$_3$Sb$_5$ and RbV$_3$Sb$_5$. Therefore, the changes in CDW intensity with magnetic fields have now been reproduced four times using STM, in two different compounds and by three different groups, in contrast to the earlier paper by the comments' authors[2], which claimed no changes with magnetic fields. Beyond STM, macroscopic properties such as the resistivity anisotropy were also shown to be induced by an external magnetic field[6].

Second, we would like to emphasize that in our approach, we focus on the relative intensities of the CDW peaks ($I_r = \frac{(I_1 - I_3)}{(I_1 + I_3)/2}$) and on the ratios of the Bragg vectors ($\boldsymbol{Q}_r = |\boldsymbol{Q}_{B1}|/|\boldsymbol{Q}_{B3}|$), rather than the individual values ($I_1$, $I_3$, $\boldsymbol{Q}_{B1}$, $\boldsymbol{Q}_{B3}$). By tracking the fidelity of this single figure-of-merit with respect to the long sequence of change in magnetic field direction and laser polarization, one can robustly and easily tell if the response of CDW peaks and Bragg vectors to these external stimuli is governed by random drift and tip change (~50% fidelity), or an intrinsic response of the sample (>90% fidelity), compared to tracking individual values.

Finally, random drift and tip changes could not possibly cause the systematic variation and high fidelity of $I_r$ and $\boldsymbol{Q}_r$ with laser illumination and magnetic fields observed in our work. Indeed, to demonstrate reproducibility and the robustness of our results, we performed more than 30 exposures to laser polarized in two different directions, and in every instance, $I_r$ and the corresponding $\boldsymbol{Q}_r$ showed consistent and well-defined changes (100% fidelity). Invoking random drift and tip changes to explain these reproducible and systematic effects would require an extreme coincidence. When combined with the fact that the same behavior was observed in two different compounds and by 3 different STM groups, as explained above, we can conclude that the random processes invoked by the authors of the comment cannot address the experimental observations.

## 2. CDW switching with magnetic field

### 2.1. Intensity of Fourier peaks

The authors of the Comment argue that random change in tip sharpness anisotropy can introduce randomness. In our data[1], $I_r$ show a systematic change when exposed to the laser (100% fidelity) and magnetic field (92% fidelity). As previously mentioned, it is impossible to produce such systematic and reproducible changes by random tip changes and drift.

### 2.2. Atomic Bragg vector length changes

Here we address the measured lattice distortion which goes hand in hand with the CDW intensity changes.

First, we note that Candelora et al. chose to discuss only the magnetic-field related effects, which contradict their previous results[2], and not the substantial light-induced Bragg vector ratio $Q_r$ changes we have reported. The latter shows a clear correlation with the direction of the electric field, something which would be impossible to explain on the basis of drift or tip changes.

Second, the authors analyze magnetic field data (backward scans) that was not part of our publication. Indeed, we had not previously analyzed this data for our paper. We agree with a specific point made by Candelora et al., that the $Q_r$ values extracted from the backward scans in scan #102,103,106 show differences from the forward scans. However, we note that these particular data sets were collected exclusively with a smaller (30nm) field of view. The smaller scan size leads to larger digitized pixels in Fourier space, reducing sensitivity in tracking small motion of Bragg peaks. This point is quantitatively demonstrated in the Candelora, et al.'s Supplementary Fig. 3. With the 40nm scan size (scan #61, #66), the change in $Q_r$ from -2T to 2T is ≈1.0%, twice larger than the error bar 0.5% (from Fig. 3c of Ref[1]). By cropping these same scans to 30 nm, now the change in $Q_r$ is significantly reduced to ≈0.25%, even smaller than the error bar (0.5%). This explains the inconsistency in the $Q_r$ between the forward and backward scans for small scans (30nm). In comparison, the larger (40nm) scan size shows consistency in the variations in $Q_r$ between the forward and backward scans. Another point here is that the range of variation for the magnetic field data is much smaller than the laser data: the $Q_r$ variation between different magnetic fields is ≈1% while the range of $Q_r$ observed in the laser experiments is ≈2.5%. *This is precisely why the laser experiments are so important, as they validate the methodology that is consistently applied to all data sets.*

The key point is that, despite the differences between forward and backward scans for the Bragg vector ratios, the response of the intensity of the CDW peaks ($I_r$) to the magnetic field is consistent with the results reported in our paper. $I_r$ is not as affected by scan size as it only integrates the weights within a pixel and is not multiplied by its position. As shown in Fig. R1a, we analyze the sign of the intensity contrast $I_r$ along the magnetic field sequence for both forward and backward scans using the same methodology as in our paper. As seen from the Fig. R1a, for the 12 data points, both scans show $I_r$>0 for B= -2T and negative $I_r$<0 for B= +2T, except for one point (scan 66 BWD). The 92% fidelity of the magnetic field induced switching of the CDW intensity order demonstrates the robustness of the conclusions of our paper.

Our data demonstrate that the CDW is extremely sensitive to strain. In fact, our experiments on strained regions (near a wrinkle) show that the CDW is pinned and does not respond to magnetic fields (Fig. R1b), which may explain the authors' previous null results[2].

### 2.3. *Scan speed and magnetic field values*

Candelora et al. raise concerns about our reported magnetic field values as well as our conclusions based on scan speeds and z-piezo values. As we elaborate on in the Reply Supplementary information, these statements are factually incorrect and should not be considered in this discussion.

### 3. CDW switching with laser illumination

Candelora et al. argue without analysis that the observation of the laser-induced manipulation of the CDW intensity is filled with artifacts. However, the 100% correlation between the sign of $I_r$ and the direction of polarization along the 33 data points in our data[1] proves that the observation is not from an artifact and demonstrates that using the CDW intensity contrast $I_r$ and ratio between Bragg vectors $\mathbf{Q}_r$ successfully removes potential effects from tip changes.

**Conclusion**

To conclude, in their comment, the authors ignore previous literature where the same effects have been seen, make unsubstantiated statements about the laser manipulation of the CDW, and disregard the systematic and reproducible changes observed by us which would require an impossible series of coincidences.

**Figure R1**

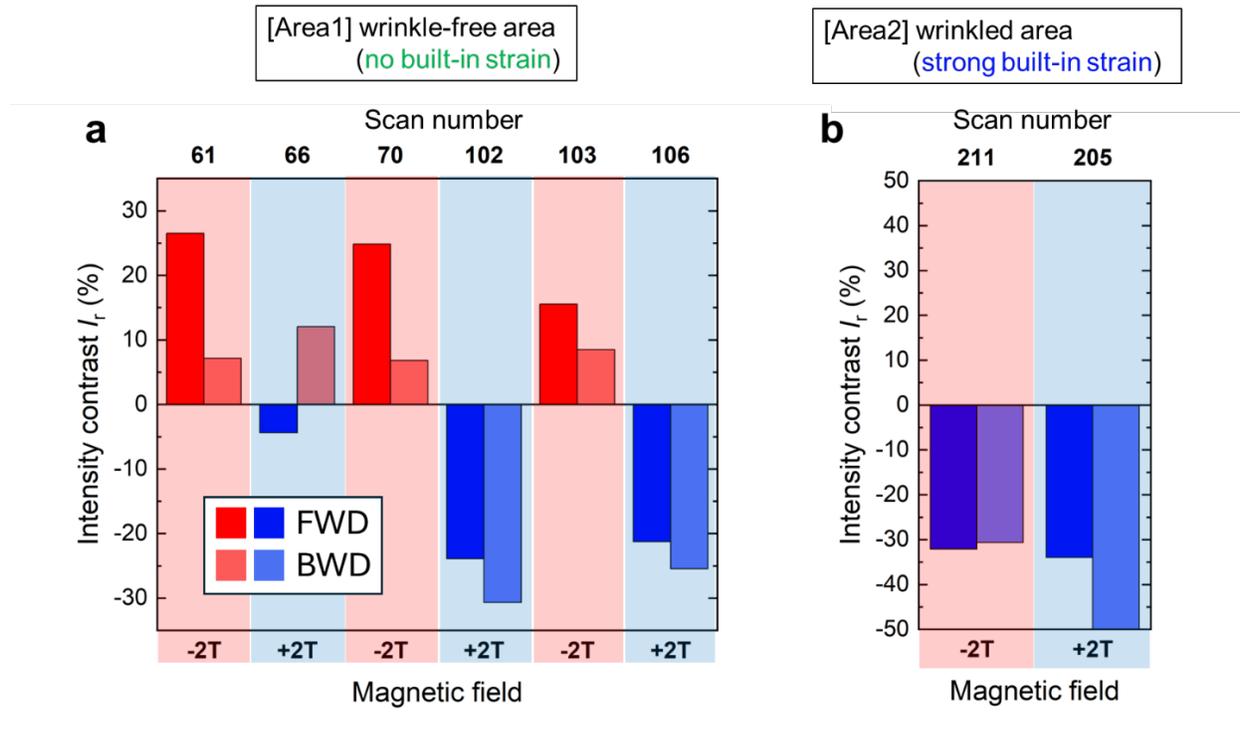

**Fig. R1: Comparison of the sign of CDW intensity contrast between FWD and BWD scans along the magnetic field sequence.**

**a**, Result from a wrinkle-free area (Fig. 4g in Ref[1]), showing a robust (92%) correlation of the sign of $I_r$ and the direction of the field throughout the FWD and BWD scans. **b**, Result from an wrinkled area (Ext. Data Fig. 9 in Ref[1]), showing the same sign of $I_r$ regardless of FWD and BWD scans, and showing no correlation between the sign of $I_r$ and the direction of the field due to strong built-in strain from the wrinkle on the neighbor.

# Supplementary Information of Reply

In the supplementary Note of the Comment from Candelora et al., the authors present arguments related to scan speed and tip height upon application of a magnetic field to suggest that the magnetic field effects we demonstrate in our paper cannot be correct.

Before we respond to this, we reiterate that our primary findings i.e., the sensitivity of the CDW Bragg peaks to magnetic fields **has been reported both before and after our work in three distinct STM publications** (Nat. Mater. 20, 1353 (2021); Phys. Rev. B 104, 035131 (2021) and Nature 632, 775 (2024)). Moreover, sensitivity of anisotropic properties of the CDW phase to magnetic fields was also seen in a transport paper (Nat. Phys. 20, 584 (2024)).

The crux of the argument from the comment rests on extrapolating the behavior of the authors' STM in a magnetic field, to all STMs from UNISOKU Co., Ltd. But this is not a logically sound extrapolation. In fact, the actual size of the small nanometer changes in the z-height upon application of a magnetic field depends sensitively on the details of the STM design as well as construction. As we will show in this supplemental reply, any change in z-piezo height upon application of a magnetic field varies greatly depending on the system and the tip. For example, the Unisoku 300mK STM used for the magnetic field studies in this paper is a custom UHV STM built into a Janis cryostat which shows different behavior from another UHV 300mK STM in our lab which is built into Cryogenic Inc. cryostat. We would also like to draw attention to the fact that many of the statements in this comment supplementary information are factually incorrect.

In general, while we welcome constructive discussion about different physical interpretations of our paper[1], it is detrimental to the community to make unjustified generalizations about the behavior of instruments.

1. **Regarding the influence of scan speeds on the CDW intensities**

First, the difference in the scan speed does not change our observed phenomena. For example, as seen in Fig. 3a of Xing et al.[1], along the laser polarization sequence, the scan speed varies between 78 to 130 nm/s which is factor of ~1.7. Along this sequence, the sign of the CDW intensity contrast shows 100% fidelity with respect to the direction of polarization. This observation clearly demonstrates the fact that the range of differences in the scan speed in our study does not affect the switching of CDW intensity contrast upon external stimuli.

Second, the assertion that we slowed down the scan speed by a factor of 2-3 between scans at B = 2 T compared to scans at B= - 2 T. This is factually wrong. Scans that show topographies at B = -2T are #61, #70, #103, and topographies at B = + 2T are #66, #102, #106. Comparing the scan speeds between data at -2T and +2T, #61(-2T) and #66(+2T) show exactly the same scan speed of 156.2 nm/s. Similarly, #70(-2T) and #102(+2T) also show exactly the same scan speed of 98.7 nm/s. The only difference is from the last set where #103(-2T) shows 98.7 nm/s and #106 shows 197.4 nm/s. Thus, we are puzzled about where the comment's assertion arises from ("-2T scans are systematically 2-3 times slower than +2T scans").

2. **Regarding the absence of magnetic-field effects on the CDW data from the authors' previous work**

The comment argues since an analysis of 300 topographies in multiple regions obtained by the Zeljkovic group [2,3] do not show the magnetic field related effects observed by us (and others), we must all be wrong.

This is not a sound argument. There are many potential reasons that may have prevented Candelora et al. from seeing the effects. One of the likely reasons, which is the presence of non-negligible residual strain in the sample, is actually consistent with our own findings reported in Fig. R1b of our Reply.

*The important point here is that science progresses by replication and our work has been replicated in three other STM papers*.

3. **Regarding the possible changes in z-height under a magnetic field**

The comment argues that the reported magnetic field values in our paper must be incorrect since the changes in the z-heights observed during the magnetic field switching (from -2T to +2T) in our measurements (~0.1 nm) are much smaller than what the authors observed in their previous publication[3] (2~3 nm).

Below, we show that one cannot extrapolate from one STM setup to another and that this assertion that the z-value can act as a proxy for magnetic field values is in fact factually incorrect.

First, from an instrument standpoint, all STM heads are hand-made and therefore not identical. Moreover, as stated earlier, the STM used in the magnetic field studies is a custom UHV STM in a Janis cryostat which is very different from the standard Unisoku system.

To further prove that the z-values are not a proxy for the magnetic field value, we conducted 3 control experiments as described below.

1) With the STM used in this study, we conducted a sequence of studies where we monitored the Z values at 1.7 K (the same temperature condition as our paper[1]) while changing the magnetic field (B). The results, shown in the format of Z (B), are: -92.5 nm (0 T) ➔ -92.1 nm (+2 T) ➔ -92.5 nm (0 T) ➔ -92.6 nm (-2 T). The difference in Z is 0.4 nm between 0 T and 2 T, which again reproduces similar value (~0.1 nm) reported in our paper[1] rather than the value of 2-3 nm from comment's authors.

2) With the same STM we conducted another magnetic field sequence after changing the tip. The results are as follows: -52.6 nm (0 T) ➔ -43.6 nm (-2 T) ➔ -35.7 nm (0 T). We find that even with the same STM, the fluctuation of Z could vary by an order of magnitude from experiment to experiment. Second, the Z also doesn't shift in a monotonic way unlike what is described in comment's authors' paper[3].

3) We also extracted the Z height in a recent magnetic field study in our group performed with a different 300mK STM with a cryostat from the same company as the comment's authors' STM (Cryogenic Inc). These values are extracted from scans taken under the same conditions, at the same location but under different magnetic fields. The result is the following: -102 nm (0 T), -101 nm (1 T), -101 nm (5 T) and -106 nm (8 T). The identical Z heights at 1 T and 5 T, and the non-monotonic Z shift from 0 T to 8 T further contradict all assertions in the comment.

Taken together, these three experiments demonstrate no consistent relationship between the Z-shift and the applied magnetic field and demonstrates that Z is not a reliable proxy for the applied field value under these conditions.

**Supplementary References**